# Thermodynamic properties of fcc lead: A scalar and fully relativistic first principle study

Balaram Thakur[1*], Xuejun Gong[1,2], and Andrea Dal Corso[1,2]

[1]International School for Advanced Studies (SISSA), Via Bonomea 265, 34136 Trieste, Italy.

[2]CNR-IOM, Via Bonomea 265, 34136 Trieste, Italy.

*Corresponding author: bthakur@sissa.it

**ABSTRACT:**

This study investigates the thermodynamic properties of face-centered cubic lead (fcc-Pb) using ab-initio methods within the quasi-harmonic approximation (QHA), examining the influence of spin-orbit coupling (SOC) and the exchange-correlation functionals. Two types of ultrasoft pseudopotential (US-PP) are considered: one that excludes (scalar relativistic PP) and one that includes the SOC effects (fully relativistic PP). Further, for each PP, we test the performance of three popular exchange-correlation functionals: Perdew-Burke-Ernzerhof generalized gradient approximation (PBE) (Perdew *et al.* Phys. Rev. Lett. 77, 3865 (1996)), PBE modified for dense solids (PBEsol) (Perdew *et al.* Phys. Rev. Lett. 100, 136406 (2008)), and local density approximation (LDA) (Perdew *et al.* Phys. Rev. B 23, 5048 (1981)). We calculate the Helmholtz free energy, incorporating lattice vibrations (phonons) and electronic excitation contributions. The estimated equation of state (at 4 K and 301 K), phonon dispersions (at 100 K and 300 K), mode-Grüneisen parameters ($\gamma_{q\eta}$) (at 100 K), volume thermal expansion coefficient ($\beta$), isobaric heat capacity ($C_P$), bulk modulus ($B_S$), and thermodynamic average Grüneisen parameter ($\gamma$) are compared with the available experimental and theoretical studies. Moreover, the 0 K pressure-dependent elastic constant-coefficient ($C_{ij}$) of fcc lead and Pugh ratio, Debye temperature, and longitudinal and transverse sound velocities for polycrystalline lead are presented. The contributions of electronic excitations in all the thermodynamic properties are found to be negligible. With increasing pressure, the role of spin-orbit effects decreases but does not vanish. Our findings demonstrate that SOC leads to results distinct from the SR approach, but agreement with the experiment is not consistently improved by including SOC.

**Keywords**- Lead, relativistic first principle calculation, thermodynamic properties, high pressure, and high temperature

Email:   Balaram Thakur (bthakur@sissa.it), Xuejun Gong (xgong@sissa.it),
         Andrea Dal Corso (dalcorso@sissa.it)

**INTRODUCTION:**

Lead (Pb), a soft, silvery-grey metal, also known as plumbum in Latin, exhibits high corrosion resistance and finds applications in various domains, including rechargeable batteries, cable sheathing, paints, and corrosion-resistant pipes [1,2]. Also, lead serves as an effective shielding material against X-ray and gamma radiation when incorporated into concrete [3] or used in block form. Beyond its practical applications, a nearly free electron system like lead provides an ideal platform for theoretically investigating the role of relativistic effect on a material's electronic structure and physical properties.

The relativistic effects directly shift and split the energy of the valence orbitals and indirectly influence the valence wavefunctions through the orthogonality to the core states. A general approach to introducing this in electronic structure calculation is modifying the pseudopotential (PPs). Theoretically, a non-relativistic Kohn and Sham equation is solved using the PPs modified to reproduce the solutions of fully relativistic atomic-Dirac-like equations [4,5]. Popularly, there are PPs where the shift in the energy levels due to the scalar relativistic (SR) effects is included and fully relativistic (FR) PPs where the spin-orbit coupling (SOC) is included in addition to the SR effects. The impact of the relativistic and SOC effects is particularly pronounced in lead. SOC, in particular, significantly splits the $6p$ states into $6p_{1/2}$ and $6p_{3/2}$, separated by approximately 1 eV [6], thereby exerting a substantial influence on lattice dynamics and crystal structure.

The phonon dispersions of lead is challenging to study accurately because of several anomalies observed in inelastic neutron scattering experiments [7] and the small interatomic force constant due to $sp$ electrons [8]. De Gironcoli [9], employing the density-functional perturbation theory (DFPT), successfully explained the anomaly in the longitudinal (L) branch from **Γ** to **X** along the Σ line (in [110] direction). However, the significant softening of the transverse (T) mode at **X** along Δ line (in [100] direction) [7] remained unexplained in this study [9]. Chen and Overhauser [10] attributed these depressions to the presence of spin-density waves. Subsequent studies, including those by Dal Corso [8] using ultrasoft pseudopotentials (US-PP), and Verstraete *et al.* [6] using norm-conserving pseudopotentials, demonstrated that the inclusion of SOC in the PP appreciably corrects the phonon modes at many **q** and helps significantly in softening the transverse (T) mode at **X**. Furthermore, considering the SOC, substantially improves the description of Kohn anomalies along the **Γ** to **X** and **Γ** to **K** directions [8].

Apart from the phonon dispersion, the SOC exerts a significant influence on the phase transformation in lead. Experimental studies [11,12] report that lead undergoes pressure-induced structural transitions in a sequence of fcc-hcp-bcc, with the former occurring at ~13 GPa and the latter at ~109 GPa. A recent theoretical study [13] confirms these observations and further demonstrates that the role of SOC effects is weak for fcc to hcp transition but becomes significant for hcp to bcc pressure-induced structural transitions.

The thermodynamic properties, such as volume thermal expansion coefficient and isobaric heat capacity, using the PAW method and LDA and GGA functionals without considering the SOC effect are discussed in Ref. [14]. Studies [13] incorporating SOC effects have been limited to the determination of volume-dependent elastic constant coefficients (at 0 K), isothermal bulk modulus ($B_T$), and other transport properties, primarily employing the Full-potential linear-muffin-tin orbital (FP-LMTO) method and PBEsol.

Within the quasi-harmonic approximation, it has become quite common to study the performance of different functionals (LDA, PBE, and PBEsol) on the thermodynamic properties of solids. LDA, while known to slightly underestimate the lattice parameters, generally provides more accurate phonon frequencies than GGA, particularly for 5-period metals. Conversely, PBE, although overestimating lattice parameters, often yields more accurate bulk moduli compared to LDA, which tends to predict too hard bulk moduli. PBEsol [15], by modifying the exchange energy part of the PBE, improves the lattice parameter of solids, at the expense of the accuracy of the functional in molecules.

The LDA, PBE, and PBEsol functionals are simple to use and although for lattice parameters, bulk moduli, and phonons one has the advantage over the other, on thermodynamics, it is not yet clear which is the best functional to use. For example, our recent studies on iridium [16] demonstrated that PBEsol provides more accurate lattice parameters, while both LDA and PBEsol exhibit good agreement with the experimental results for phonon dispersions, isobaric heat capacity, and low-temperature thermal expansion calculation, while the PBE functional is better for high-temperature thermal expansion and bulk moduli. However, in rhodium [17], we found that PBEsol showed better agreement for volume thermal expansion at low temperatures, whereas PBE performed better at higher temperatures. Additionally, PBE appeared to provide more accurate heat capacity predictions within the studied temperature range.

A recent study on thorium ($Th^{90}$) by Kývala and Legut [18] using PBE and PBEsol found that SOC does not affect thermodynamic properties. However, these observations may not hold in other heavy metals. Given the potential significance of both *xc* functional choice and SOC effects on the thermodynamic properties of heavy metals, this study aims to address this knowledge gap. Actually, such studies are computationally expensive, particularly for heavy elements where considering the SOC effect becomes mandatory. To our knowledge, a comprehensive study of this nature has not been previously conducted for lead.

Therefore, in the present study, we focussed on three popular *xc* functionals: LDA, PBE, and PBEsol, and studied their performance when used with fully and scalar relativistic pseudopotentials to calculate the thermodynamic properties of lead. The contribution of phonons and electronic excitation to Helmholtz free energy is considered. The equation of state at 4 K and 301 K, phonon dispersion curves at 100 K and 300 K, and mode-Grüneisen parameters ($\gamma_{q\eta}$) (at 100 K) are discussed. The temperature and pressure-dependent volume thermal expansion coefficient ($\beta$), isobaric heat capacity ($C_P$), bulk modulus ($B_S$ and $B_T$), and thermodynamic average Grüneisen parameter ($\gamma$) are illustrated. Finally, we discuss the 0 K pressure-dependent elastic constant-coefficient of fcc-lead and Pugh-ratio of polycrystalline lead.

**METHODS:**

The thermodynamic properties of lead were calculated using the density functional theory (DFT) within quasi-harmonic approximation (QHA). The calculations were performed using the `Thermo_pw` [19] code, a driver of the Quantum ESPRESSO (QE) routines [20,21]. We employed scalar relativistic (SR) and fully relativistic (FR) ultrasoft (US) [22] PPs (from *pslibrary* [23,24]). These US-PPs were generated according to a modified Rappe-Rabe-Kaxiras-Joannopoulos ultrasoft (RRKJ) [25] scheme. The formalism of including the SOC in ultrasoft PPs (USPP) [22] is discussed in detail in Refs. [5,26].

For each SR and FR PPs, the exchange-correlation (*xc*) functional is approximated by the local-density approximations (LDA) with the Perdew-Zunger [27] parameterization, the generalized gradient approximation (GGA) suggested by Perdew-Burke-Ernzerhof (PBE) [28], and the PBE functional modified for densely packed solids (PBEsol) [15]. For LDA, PBE, and PBEsol, we used `Pb.pz-dn-rrkjus_psl.0.2.2.UPF`, `Pb.pbe-dn-rrkjus_psl.0.2.2.UPF` and `Pb.pbesol-dn-rrkjus_psl.0.2.2.UPF` file for SR PPs, whereas `Pb.rel-pz-rrkjus_psl.0.2.2.UPF`, `Pb.rel-pbe-dn-rrkjus_psl.0.2.2.UPF` and `Pb.rel-pbesol-dn-rrkjus_psl.0.2.2.UPF` file for FR PPs, from *pslibrary* [23,24]. The electronic configuration of lead is [Xe] $4f^{14}5d^{10}6s^26p^2$. In the above PPs, the 5*d*, 6*s*, and 6*p* were treated as valence states, while 4*f* states are frozen in the core and are accounted for by the nonlinear core correction [29]. For clarity, the calculations performed using SR US-PPs with LDA, PBEsol, and PBE are hereafter referred to as LDA[SR], PBEsol[SR], and PBE[SR], respectively. Similarly for FR US-PPs, the designations LDA[FR], PBEsol[FR], and PBE[FR] are used.

The pseudo-wavefunctions (charge densities) are expanded in a plane wave basis with a kinetic energy cutoff of 60 Ry (350 Ry). For the Brillouin zone (BZ) integration, a uniform 32 × 32 × 32 **k**-point mesh using the Monkhorst-Pack method [30] is utilized. The presence of the Fermi surface is dealt with by the Methfessel and Paxton (MP) smearing approach [31] with a MP smearing parameter $\sigma = 0.02$ Ry. Density functional perturbation theory (DFPT) [32] extended to ultrasoft PP [26] is used to calculate the dynamical matrices on a coarse **q**-point grid. Subsequently, these dynamical matrices were interpolated onto a thicker 192 × 192 × 192 **q**-point grid using the Fourier interpolation method. For the SR case, a 10 × 10 × 10 **q**-point grid is used for LDA, PBE, and PBEsol functionals. For the FR case, a **q**-point grid of 10 × 10 × 10 for PBE and an 8 × 8 × 8 for LDA and PBEsol was found sufficient to satisfactorily account for the presence of several anomalies in phonon dispersion of lead.

For the SR case and all three functionals, the dynamical matrices are calculated on 9 geometries with a step of 0.1 a.u. (LDA: 8.7 to 9.5 a.u., PBEsol: 8.8 to 9.6 a.u., and PBE: 9.0 to 9.8 a.u.). For the FR calculations, the following geometries were used: LDA: 7 geometries (8.8 to 9.4 a.u. in steps of 0.1 a.u.), PBEsol: 9 geometries (8.8 to 9.6 a.u. in steps of 0.1 a.u.), and PBE: 11 geometries (9.25 to 9.75 a.u. in steps of 0.05 a.u.). This choice ensures that we are in regions of parameters where the phonon modes are real and positive across all the studied geometries.

Finally, the Helmholtz free energy is calculated using the phonon density of states obtained for the different volumes. In Helmholtz free energy, the contribution of electronic excitation is also included within the rigid band's approximation. The formulas used for evaluating different thermodynamic properties and the methodology applied for various metals are documented in the `Thermo_pw` [19] manuals and in [16,17,33–37]. For single crystals elastic constant coefficients ($C_{ij}$), the stress tensor components were determined by applying the strain on the equilibrium geometry at 0 K. Furthermore, using this single crystal $C_{ij}$'s values and the Voigt-Reuss-Hill (VRH) averaging relations [38], the polycrystalline shear modulus (G), bulk modulus (B), compressional and shear velocities, Pugh ratio (G/B), and Debye temperature ($\theta_D$) are evaluated.

**RESULTS AND DISCUSSION:**

In Figure 1, we compare the pressure-volume (P-V) equations of state (EOS) obtained using the FR (solid lines) and SR (dashed lines) US-PPs for three different functionals obtained at 301 K and 4 K (in inset). The effect of SOC is most pronounced only at low pressure, while the differences in volume between FR and SR calculations decrease with increasing pressure. Comparing our results with the available experimental reports, we find that the room-temperature data from [11] (square) and [39] (circle) lies in the midway of our LDA and PBEsol isotherms at 301 K. The results of Mao *et al.* (triangle) [12] at ~ 4.4 GPa and 6.5 GPa are close to the LDA result, while they follow the PBEsol at higher pressures. At low temperatures (80 K), the experimental values of Strässle *et al.* [39] (circle) lie in the midway of our LDA and PBEsol result at 4 K.

The theoretical reports of Strässle *et al.* [39] determined using the PAW method with SOC at 0 K for LDA and PBEsol agrees with our results (see inset of Figure 1). The values reported by Söderlind and Young [40] at 0 K, calculated using the all-electron FP-LMTO method and the GGA of PW91 (SOC effect for *d* and *f* states, not for *p* state) are lower than our PBE results (inset of Figure 1). This discrepancy can be attributed to the difference between the equilibrium volumes: 214.5 a.u.$^3$ in their study compared to ours 217.2 a.u.$^3$ (for PBE$^{FR}$) at 0 K and 0 GPa. Additionally, the room-temperature

Smirnov [13] values obtained using FP-LMTO and PBEsol$^{FR}$ follow reasonably with our PBEsol result in Figure 1.

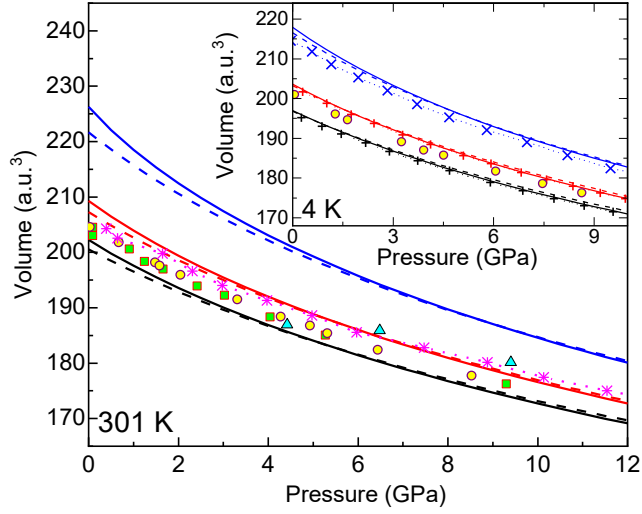

**Figure 1.** P-V equation of state (EOS) of lead calculated at 301 K and 4 K (in inset) for FR (solid lines) and SR (dashed lines) US-PPs using LDA (black), PBEsol (red), and PBE (blue) functionals. The experimental results of Kuznetsov *et al.* [11] (square) (300 K, 204.51 a.u.$^3$), Mao *et al.* [12] (triangle) (300 K, 204.7 a.u.$^3$), Strässle *et al.* [39] (circle) (298 K, 204.5 a.u.$^3$ in the main figure and 80 K, 200.9 a.u.$^3$ in the inset) are shown. The theoretical reports of Smirnov [13] (magenta dot and $*$) (FP-LMTO, PBEsol$^{FR}$, 300 K, 204.3 a.u.$^3$), Strässle *et al.* [39] (black dot and + for PAW, LDA$^{FR}$ at 0 K, 196.339 a.u.$^3$; red dot and + for PAW, PBEsol$^{FR}$ at 0 K, 203.095 a.u.$^3$), Söderlind and Young [40] (blue dot and $\times$) (all-electron FP-LMTO, GGA of PW91$^{FR}$, 0 K, 214.5 a.u.$^3$) are included.

In Table 1, we summarize the equilibrium unit-cell lattice constant ($a_0$), bulk modulus ($B_0$), and its pressure derivative ($B'_0$) at 0 K and 301 K, obtained by fitting the Helmholtz free energy to the 4$^{th}$-order Birch-Murnaghan (BM) equation. Here, we included the contribution from the zero-point vibrations in the Helmholtz free energy. From Table 1, our analysis reveals that the effect of SOC on the equilibrium parameter is more pronounced at higher temperatures, particularly for the PBE. For example, the difference in the lattice constant's value with and without SOC at 300 K is $\sim$ 0.3 % for LDA and PBEsol, reaching a maximum of $\sim$ 0.7 % for PBE. Furthermore, a notable decrease in bulk modulus is noticed when SOC is considered (about $\sim$ 13 %, 15 %, and 22 % for LDA, PBEsol, and PBE, respectively). This suggests a significant enhancement in the thermal expansion coefficient, particularly for PBE, including the SOC interaction, as we will discuss later.

The experimental $a_0$ determined at 5 K [14,41] agrees with our LDA (error $\sim$ 0.5 %) and PBEsol ($\sim$ 0.6 %), whereas at 301 K our LDA$^{FR}$ and PBEsol$^{SR}$ are very close ($\sim$0.4 %) to the experimental value at 291 K [42]. Similarly, experimental $B_0$ values agree with our PBEsol$^{SR}$, while our SR $B'_0$ for all the functional agrees with the experimental reports, as shown in Table 1. Comparing our work with previous theoretical studies at 0 K, the all-electron equilibrium parameters for LDA$^{SR}$ from Ref. [14] agree reasonably with our LDA$^{SR}$ values. In contrast, we notice slight discrepancies between the reported [14] PBE (PAW), and our PBE values for the FR case. Notably, our LDA$^{SR}$ and LDA$^{FR}$ results agree with the Dal Corso [8] results determined using US-PP LDA$^{SR}$ and LDA$^{FR}$. At 300 K, the a$_0$ and B$_0$ values using PBEsol$^{SR}$ with the FP-LMTO method [43], are similar to our PBEsol$^{SR}$ values. However, the reported PBEsol$^{FR}$ a$_0$ (B$_0$) in Ref. [43] are smaller (larger), respectively, than our PBEsol$^{FR}$ result.

| Exchange – Correlation Functional | | Lattice constant ($a_{To}$) (a.u.) | | Bulk modulus ($B_T$) (GPa) | | Pressure derivative of $B_T$ ($B'_T$) | |
|---|---|---|---|---|---|---|---|
| | | 0 K | 301 K | 0 K | 301 K | 0 K | 301 K |
| This study | LDA | 9.225 / 9.227* | 9.292 / 9.318* | 52.8 / 50.3* | 46.1 / 39.9* | 4.9 / 4.9* | 5.6 / 6.5* |
| | PBEsol | 9.325 / 9.329* | 9.394 / 9.424* | 47.7 / 45.2* | 41.7 / 35.5* | 4.9 / 5.0* | 5.5 / 6.7* |
| | PBE | 9.524 / 9.542* | 9.607 / 9.671* | 39.9 / 36.7* | 33.6 / 26.1* | 4.8 / 5.0* | 5.8 / 6.8* |
| Other calculations | LDA | 9.231$^a$, 9.219$^b$, 9.211*$^c$, 9.23$^e$, 9.23*$^e$, 9.222$^f$, 9.226*$^f$ | | 51.0$^a$, 52.0$^b$, 52.0$^e$, 49.6*$^e$, 52.89$^f$, 50.53*$^f$ | | 4.7$^a$, 5.0$^b$, 4.92$^f$, 4.99*$^f$ | |
| | PBEsol | 9.318*$^c$, 9.387$^d$, 9.349*$^d$, 9.325$^f$, 9.331*$^f$ | | 40.9$^d$, 41.4*$^d$, 47.95$^f$, 45.65*$^f$ | | 4.95$^f$, 5.05*$^f$ | |
| | PBE | 9.547$^a$, 9.513$^b$, 9.539*$^c$ | | 39.0$^a$, 40$^b$ | | 4.2$^a$, 5.6$^b$ | |
| Experimental studies | | 9.269 (5 K) [14,41], 9.353 (291 K) [42] | | 49.0 (low T) [14,44], 46.7 (80 K) (3$^{rd}$ BM-EOS) [39], 41.0 (298 K) (3$^{rd}$ BM-EOS) [39] | | 5.5 (0 K) [14,45], 5.5 (80 K) (3$^{rd}$ BM-EOS) [39], 5.7 (298 K) (3$^{rd}$ BM-EOS) [39] | |

**Table 1.** Equilibrium structural parameters: unit cell lattice constant ($a_{To}$), bulk modulus ($B_T$), and its pressure derivative ($B'_T$) at 0 K and 301 K for LDA, PBEsol, and PBE. Here, the values marked with and without an asterisk are measured using fully (with SOC) and scalar (no SOC) relativistic PPs. The methods are: **a**-all-electron (0 K) [14]; **b**-Projector-augmented wave (PAW) (0 K) [14]; **c**-Full-potential linearized augmented plane-wave and local orbitals method (FPLAPW+lo) (0 K) [46]; **d**-Full-potential linear-muffin-tin orbital (FP-LMTO) (300 K) [43]; **e**-Ultrasoft pseudopotential (US-PP) (0 K) [8]; **f**–PAW (0 K) [39]. The values underlined are used for comparison (see text).

Figure 2a illustrates the temperature dependence of equilibrium volume $V_{eq}$(T) for each case, obtained by fitting the free energy and interpolating in the entire temperature range. At low temperatures, the experimental results from Refs. [47,48] lie between LDA and PBEsol for SR and FR cases. However, due to the significant enhancement in the thermal expansion for the FR case, the experimental data at higher temperatures appears to be close to LDA$^{FR}$. This enhancement in the thermal expansion is pronounced in the temperature-dependent thermal expansion coefficient (β) obtained from the derivative of $V_{eq}$(T), as depicted in Figure 2b.

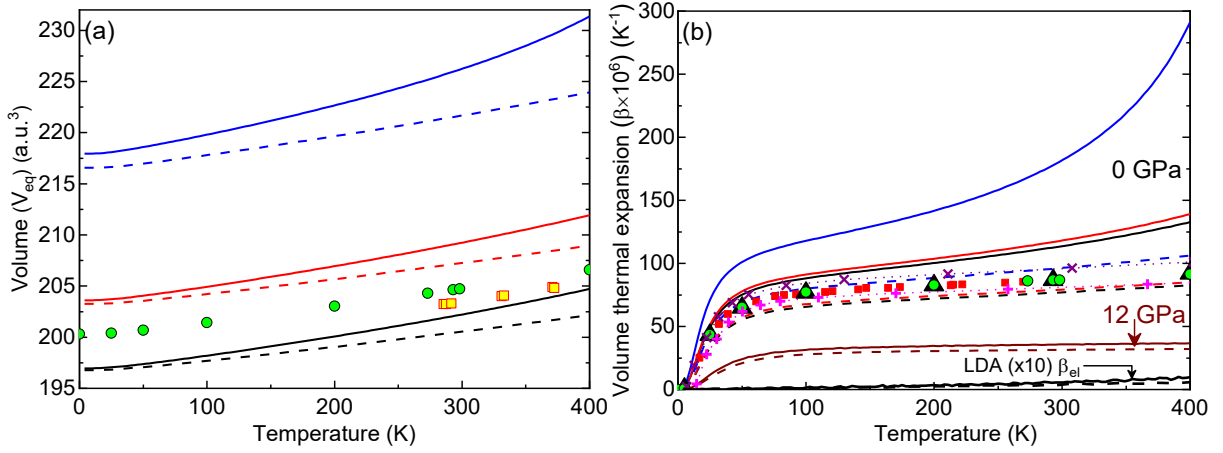

**Figure 2.** Temperature dependence of (a) equilibrium volume $V_{eq}$ (T) and (b) volume thermal expansion coefficient β(T). The solid and broken line represents the curve obtained with and without SOC, and the different *xc* functionals are LDA (black), PBEsol (red), and PBE (blue). In (a), experimental results are from [48] (circle) and [47] (square). In (b), experimental data are used from [49] (square), [41] (triangle), and [48] (circle), and calculations are from [14] (broken and + for LDA$^{SR}$) and (broken and × for PBE$^{SR}$). The variation of β with the temperature at 12 GPa (brown) and the electronic excitation contribution (×10 $β_{el}$) at 0 GPa for LDA is included.

Figure 2b demonstrates a significant effect of SOC on the thermal expansion coefficient (β) where, for all the functionals, the β for the FR case is greater than the SR case. This difference between the PPs decreases with increasing pressure but does not vanish completely, as shown in Figure 2b for LDA. Moreover, the electronic excitation contribution (EEC) to the β ($β_{el}$) for LDA in Figure 2b (scaled up tenfold) indicates that the phonon contribution to the free energy is much higher than the EEC. Compared with the experimental reports [41,48,49], the PBE$^{SR}$ is closer at low temperatures, whereas both LDA$^{SR}$ and PBEsol$^{SR}$ are satisfactory at higher temperatures. Our LDA$^{SR}$ and PBE$^{SR}$ agree with the SR calculation of Grabowski *et al.* [14] obtained using PAW methods. The FR β(T) is instead significantly higher than the experimental values.

Figure 3a demonstrates the temperature-dependent isobaric heat capacity ($C_P$). At 0 GPa, $C_P$(T) for LDA and PBEsol are similar for both FR and SR cases, while PBE is always higher. The anharmonic contribution at higher temperatures is due to the consequence of larger β [since $C_P$ (p, T) = $C_V$ (V, T) + $β^2$(p, T)·$B_T$(V, T)·$V_{eq}$(p, T)·T], which is much higher for PBE$^{FR}$. The best fit of the experimental results of Arblaster [50] lies between our LDA$^{SR}$ and PBEsol$^{SR}$. In contrast, the LDA$^{FR}$ and PBEsol$^{FR}$ follow the experiment reasonably well at low temperatures, whereas at higher temperatures, they are higher than the experimental data. Our PBE$^{SR}$ and LDA$^{SR}$ heat capacities and the electronic excitation contributions to $C_P$ for PBE$^{SR}$ agree with the LDA and PBE SR calculation of Grabowski *et al.* [14], as shown in Figure 3a. Notably, figure 3a illustrates that for LDA, the effect of SOC on $C_P$ diminishes significantly with increasing the pressure to 12 GPa.

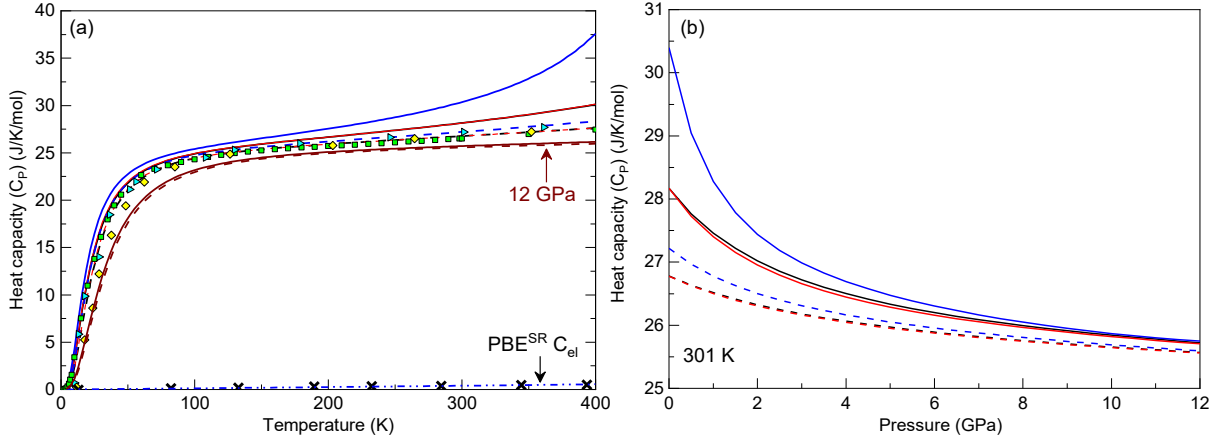

**Figure 3.** (a) Temperature and (b) pressure-dependent isobaric heat capacity ($C_P$) of fcc-lead. The black, red, and blue lines represent LDA, PBEsol, and PBE at 0 GPa. In (a), the brown line represents $C_P(T)$ for LDA at 12 GPa. The solid and broken lines are for FR and SR US-PPs. The experimental result of Arblaster (square) [50] and the theoretical calculation of Grabowski *et al.* [14] using the PAW method and PBE$^{SR}$ (triangle) and LDA$^{SR}$ (diamond) are included. The electronic excitations contribution to the isobaric heat capacity ($C_{el}$) using the PBE$^{SR}$ calculation of Grabowski *et al.* [14] (×) is compared with the present work $C_{el}$ obtained with PBE$^{SR}$.

Figure 3b depicts the pressure dependence of $C_P$ at 301 K. Our results in Figure 3b show that with increasing the pressure, the $C_P$ obtained for different functionals converges, and the difference in the values obtained from FR and SR calculations diminishes. At significantly high pressure, the anharmonic term in the $C_P$ (mentioned above) will diminish, and $C_P$ will approach the Dulong-Petit limit of $3k_B$ (~ 25 J/K/mol).

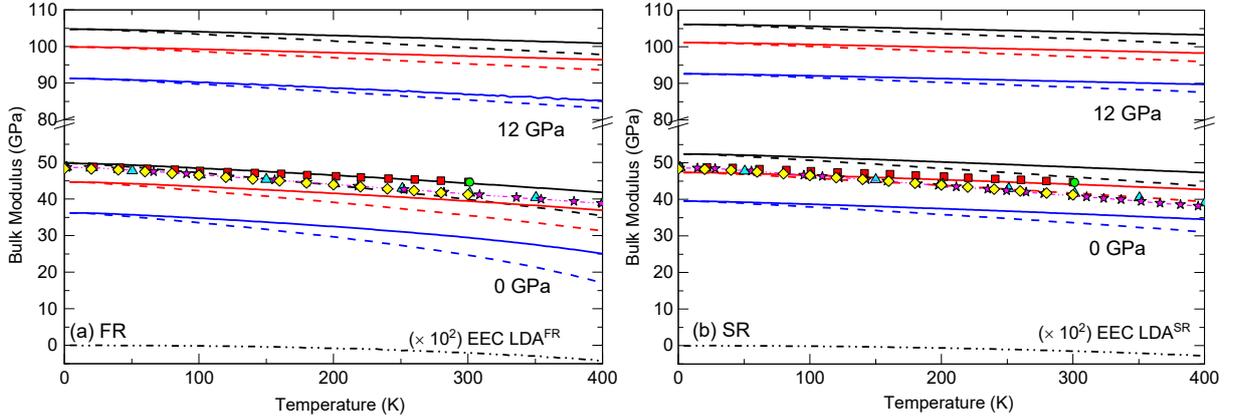

**Figure 4.** Temperature-dependent isentropic ($B_S$) (solid lines) and isothermal ($B_T$) (dashed) bulk modulus at 0 GPa and 12 GPa using (a) FR and (b) SR US-PPs. The variations for LDA, PBEsol, and PBE functionals are given in black, red, and blue. The EEC to the $B_T$ for LDA$^{FR}$ and LDA$^{SR}$ (scaled up $10^2$ times) at 0 GPa is shown in (a) and (b), respectively. Symbols are experimental results from Waldorf ($B_S$) (from elastic moduli) (square) [51], Vold ($B_S$) (circle) [52], Cordoba ($B_T$) (triangle) [53], and Strässle (diamond) (determined from experimental EOS) [39]. The theoretical result ($B_T$) of Smirnov [13] (dot and star) using PBEsol with and without SOC is included in (a) and (b), respectively.

Figure 4 presents the temperature-dependent isoentropic ($B_S$) and isothermal ($B_T$) bulk modulus at 0 GPa and 12 GPa. Similar to the thermal expansion coefficient (β), for all the cases, we observe negligible EEC to the $B_T$, and for LDA it is shown in Figure 4. The experimental results (both $B_S$ and $B_T$) from [39,51–54] agree with the LDA$^{FR}$ in Figure 4(a) and PBEsol$^{SR}$ in Figure 4(b). The $B_T$ of Smirnov [13] with SOC are higher than our PBEsol$^{FR}$ results (in Figure 4a), whereas, without SOC interaction, the PBEsol results of Smirnov [13] are in good agreement with our PBEsol$^{SR}$ (see Figure 4b). The difference between $B_S$ and $B_T$, which is given by $\frac{\beta^2(p,T) \cdot B_T^2(V,T) \cdot V_{eq}(p,T) \cdot T}{C_V(V,T)}$, is maximum for PBE due to the significant contribution of β (see Figure 2b). This difference for PBE at 0 GPa pressure is ~ 8 GPa and ~ 3 GPa for FR and SR cases, respectively, whereas the corresponding values remain at ~ 2 GPa when the pressure is increased to 12 GPa.

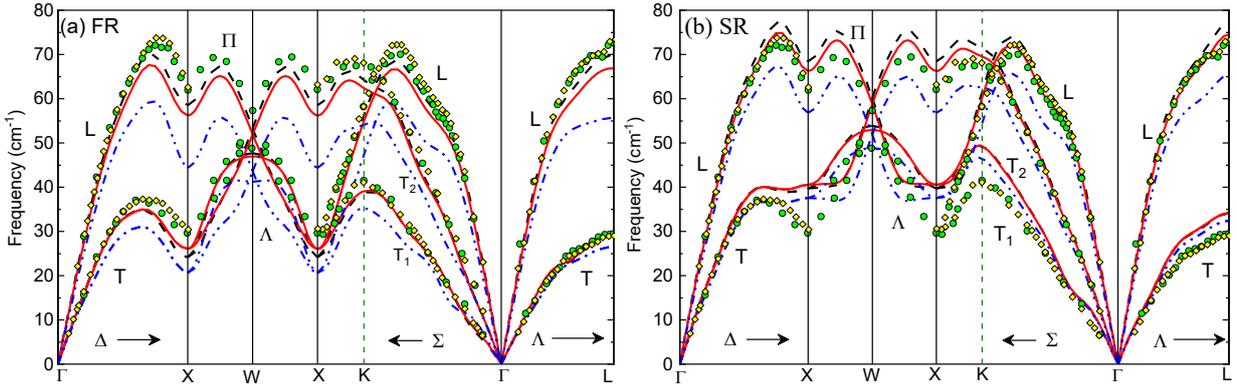

**Figure 5.** Phonon dispersion for the fcc-lead at 100 K using (a) FR (b) SR US-PPs for LDA (black broken lines), PBEsol (red solid lines), and PBE (blue dash dot-dot), as a function of reduced wave vectors along the principal symmetry directions Δ, Σ, and Λ. The phonon dispersion curve is interpolated at the lattice constant corresponding to T = 100 K. For the FR case, the unit cell lattice constant at 100 K is 9.255 a.u. (LDA), 9.358 a.u. (PBEsol), and 9.579 a.u. (PBE) whereas for SR, it is 9.247 a.u. (LDA), 9.347 a.u. (PBEsol) and 9.551 a.u. (PBE). The symbols circle and diamond are experimental results measured at 100 K [7] and 80 K [55], respectively.

Figure 5 illustrates the phonon dispersions interpolated for the lattice constant at 100 K, obtained for LDA, PBEsol, and PBE using (a) with and (b) without SOC effects. Notably, we observe that for all the functionals, the phonon frequencies are higher when SOC effects are not considered. Furthermore, the characteristic Kohn anomalies of the L branch [8] along Δ (close to **X**), Σ (from **Γ** to **K** to **X**), and Λ (close to **L**) are reproduced in all cases. Dal Corso [8] found that these anomalies depend on the sharpness of the Fermi surface, and appear for small values of smearing parameter σ. Our LDA$^{FR}$ results show good agreement with those of Dal Corso [8] obtained at 9.23 a.u. In addition, our PBE$^{SR}$ and LDA$^{SR}$ results are consistent with the Grabowski *et al.* [14] study, where the phonon dispersion at 100 K was determined at 9.530 a.u. and 9.229 a.u for PBE$^{SR}$ and LDA$^{SR}$, respectively. In comparison with the inelastic neutron scattering measurement at 100 K [7], we observe that with SR PP, the phonon mode dip in the T-branch at **X** is not replicated well. In contrast, a significant softening of this mode is found in FR PP, consistent with previous calculations [8,13].

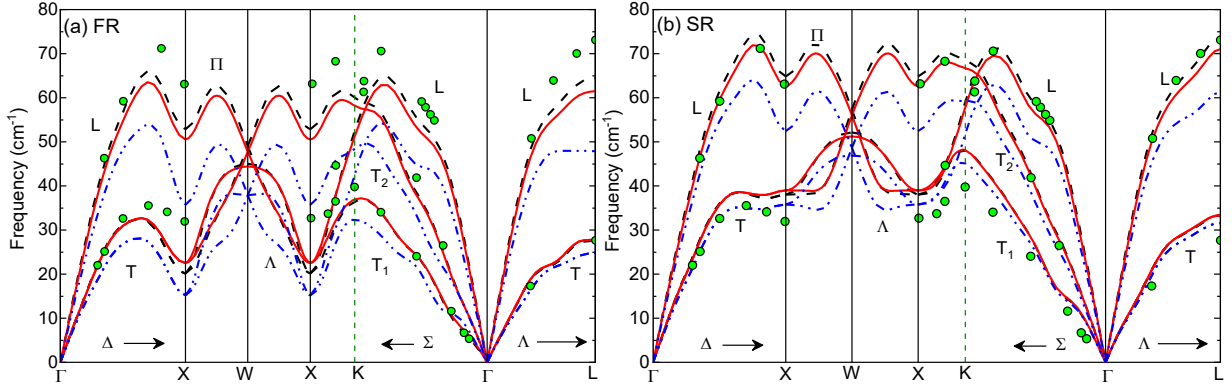

**Figure 6.** Phonon dispersion for the fcc-lead at 300 K. For the FR case, the unit cell lattice constant at 300 K is 9.317 a.u. (LDA), 9.424 a.u. (PBEsol), and 9.670 a.u. (PBE) whereas for SR, it is 9.292 a.u. (LDA), 9.394 a.u. (PBEsol) and 9.607 a.u. (PBE). The symbol circle is the data from a three-axis neutron spectrometer at 300 K from [55]. Other details are the same as in Figure 5.

Therefore, from Figure 5a, we conclude that at 100 K, LDA$^{FR}$ is the best-performing *xc* functional, whereas PBE is the worst. However, at 300 K, the phonon dispersion in Figure 6 indicates that phonon dispersion for FR PPs is significantly lower, particularly apart from the dips at **X**, PBEsol$^{SR}$ agrees well with the L branches of experimental data [55].

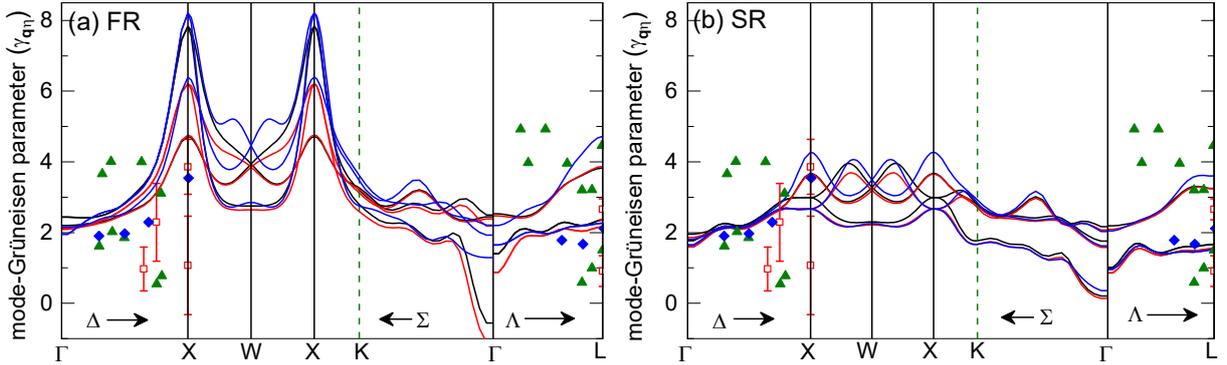

**Figure 7.** Mode-Grüneisen parameter ($\gamma_{\mathbf{q}\eta}$) obtained for different functionals using (a) FR and (b) SR PPs at 100 K. LDA, PBEsol, and PBE are represented with black, red, and blue lines. The symbols are the experimental data derived from [56] and reference therein.

Figure 7 shows the mode-Grüneisen parameter ($\gamma_{\mathbf{q}\eta}$) at 100 K along the same high-symmetry path used for phonon calculations. The experimental $\gamma_{\mathbf{q}\eta}$ data from Ref. [56] is also included for comparison. Similar to the phonon dispersion, the $\gamma_{\mathbf{q}\eta}$ values of LDA and PBEsol are similar for both FR and SR cases, but PBE consistently is higher. A significant enhancement in the magnitude of $\gamma_{\mathbf{q}\eta}$ is observed for the FR case at **X**. The discontinuity at **Γ** is attributed to the non-analyticity of $\gamma_{\mathbf{q}\eta}$. In Figure 7a, at **Γ** (along [110]), the $\gamma_{\mathbf{q}\eta}$ for PBEsol$^{FR}$ and LDA$^{FR}$ are negative, whereas for PBE$^{FR}$ it is positive. The appearance of negative $\gamma_{\mathbf{q}\eta}$ is due to the choice of **q**-grid used for phonon calculation in QHA. We found that while passing from the **q**-grid of 8 × 8 × 8 to 10 × 10 × 10, the magnitude of $\gamma_{\mathbf{q}\eta}$ at **X** remained the same, and the negative mode of $\gamma_{\mathbf{q}\eta}$ at **Γ** (along [110]) disappeared.

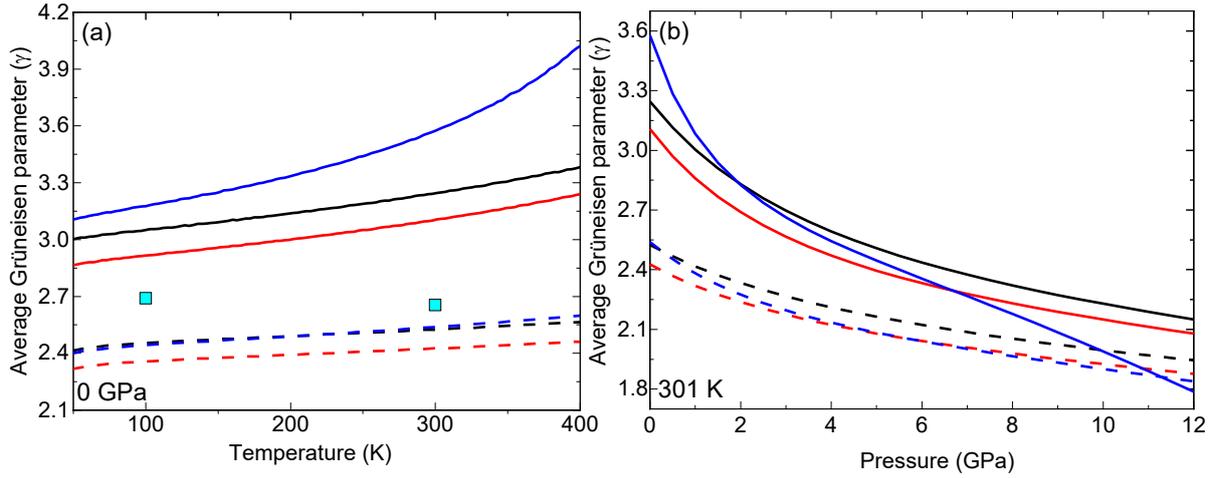

**Figure 8.** Variation of thermodynamic average Grüneisen parameter (γ) with (a) temperature at 0 GPa and (b) pressure at 301 K. The solid and broken lines are for FR and SR PPs. The LDA, PBEsol, and PBE are shown in black, red, and blue. Symbols in (a) are the experimental results [57].

| Exchange – Correlation Functional | | Elastic constant coefficient $C_{ij}$ (GPa) | | | Shear modulus $C'$ (GPa) | Approximate Debye temperature ($\theta_D$) (K) | Polycrystalline compressional sound velocity ($V_P$) (km/s) | Polycrystalline shear sound velocity ($V_G$) (km/s) |
| --- | --- | --- | --- | --- | --- | --- | --- | --- |
| | | $C_{11}$ | $C_{12}$ | $C_{44}$ | | | | |
| This study | LDA | 61  57* | 49  48* | 21  19* | 6  5* | 114.3  105.5* | 2.4  2.3* | 1.04  0.96* |
| | PBEsol | 56  52* | 43  42* | 20  18* | 6.5  5* | 113.3  104.6* | 2.4  2.3* | 1.05  0.97* |
| | PBE | 48  43* | 36  34* | 17  14* | 6  4.5* | 107.5  97.7* | 2.3  2.1* | 1.02  0.93* |
| Experiment (0K) [51] | | 55.54 | 45.42 | 19.42 | 5.06 | 105.3 | 2.3 | 1.1 |
| Calculation (PBEsol) [13] | | -  54.5* | -  43.3* | 22.25  19.98* | 5.75  5.59* | -  110.7* | -  2.3* | -  1.02* |

**Table 2:** The equilibrium elastic constant coefficient ($C_{ij}$ in GPa) at 0 K for LDA, PBEsol, and PBE. The values marked with and without an asterisk are measured using fully (with SOC) and scalar (no SOC) relativistic effects in PPs. The shear modulus ($C'$) is calculated using ($C_{11}$-$C_{12}$)/2. The approximate Debye temperature ($\theta_D$ in K), compressional (longitudinal) $V_P$, and shear (transverse) $V_G$ sound velocities (in km/s) of the polycrystalline system are calculated using the VRH averaging relations. Experimental results [51] and theoretical calculations (FP-LMTO and PBEsol) [13] are included. To determine $\theta_D$, $V_P$, and $V_G$ for Refs. [51] and [13], the experimental specific density (11.601 g/cm$^3$) at 0 K is used.

Figure 8a illustrates the temperature dependence of the thermodynamic average Grüneisen parameter (γ) for the three functionals. The results indicate that FR γ is always higher than the corresponding SR γ. The γ for LDA$^{SR}$ and PBEsol$^{SR}$ are similar. In contrast, a significant difference is observed using FR PPs. With increasing temperature, the γ for the FR PPs rises, whereas it rises less for the SR case. The experimental γ [57] at 100 K and 300 K remains between our FR and SR results. The pressure-dependent γ at 301 K in Figure 8b illustrates that with increasing the pressure, the γ decreases. The change in γ with pressure is significant in FR cases.

Finally, the equilibrium elastic constant coefficient ($C_{ij}$) obtained for LDA, PBEsol, and PBE functional with and without SOC is presented in Table 2. The result indicates that the $C_{ij}$ values for FR are generally lower than the SR counterpart. Our $C_{ij}$'s and shear modulus ($C′$) deduced from $C_{ij}$ for LDA$^{FR}$ and PBEsol$^{FR}$ are very close to the 0 K experimental values [51]. Additionally, our PBEsol results show satisfactory agreement with those of Smirnov [13]. Furthermore, using the single crystal $C_{ij}$ and the Voigt–Reuss–Hill (VRH) averaging [38] relations, approximate Debye temperature ($θ_D$), polycrystalline compressional (longitudinal) $V_P$, and shear (transverse) $V_G$ sound velocities are calculated and included in Table 2. Table 2 shows that the experimental $θ_D$, $V_P$, and $V_G$ are very close to our LDA$^{FR}$ and PBEsol$^{FR}$, whereas theoretical [13] $θ_D$ are significantly larger than our PBEsol$^{FR}$ and experimental values.

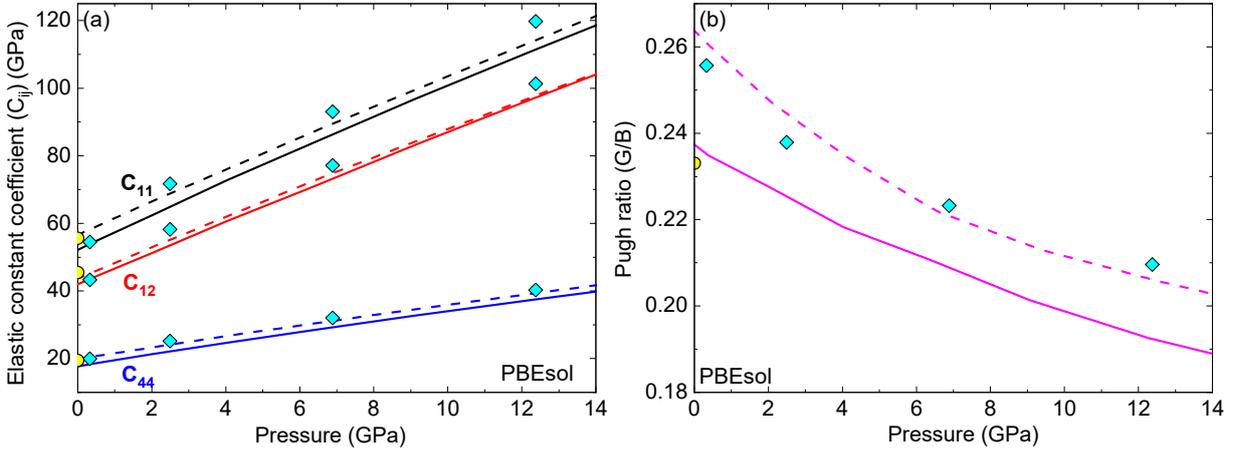

**Figure 9.** Pressure-dependent (a) elastic constant coefficient ($C_{ij}$) and (b) Pugh ratio for FR (solid line) and SR (broken lines) PPs using PBEsol. The black, red, and blue lines in (a) indicate $C_{11}$, $C_{12}$, and $C_{44}$. The symbol, circle, and diamond are the experimental [51] and theoretical (FP-LMTO and PBEsol with SOC) results [13], respectively. The Pugh ratio (G/B), a ratio of polycrystalline shear modulus (G) and bulk modulus (B), is obtained using $C_{ij}$ and VRH relations.

Figure 9a illustrates the pressure dependence of the elastic constant at 0 K for all three functionals calculated with and without considering SOC effects. The results obtained for PBEsol$^{FR}$ (solid line) and PBEsol$^{SR}$ (broken line) are shown explicitly in Figure 9a. With increasing pressure, the $C_{ij}$ value increases, and the values for the SR case always remain higher than the FR case. Furthermore, in the entire pressure range, the Born stability criteria ($C_{11} > 0$, $C_{11}$–$C_{12} > 0$, $C_{11}+2C_{12} > 0$, and $C_{44} > 0$) are satisfied, confirming all the studied structures used in QHA are stable. A similar trend is noticed for LDA and PBE (not shown here).

The $C_{ij}$'s obtained from the LMTO method and PBEsol using SOC [13] (Diamond) agreed with our PBEsol$^{FR}$ near equilibrium, whereas at higher pressures, $C_{11}$ and $C_{12}$ are higher, and $C_{44}$ agrees with our result. For comparison, we included the equilibrium experimental (circle) $C_{ij}$ at 0 K [51] (mentioned in Table 2) in Figure 9a.

Figure 9b compares the pressure-dependent Pugh ratio for our PBEsol calculations with the experimental data [51] and other PBEsol calculations [13] with SOC. The Pugh ratio (G/B) is obtained from the polycrystalline bulk modulus (B) and shear modulus (G) using the single crystals $C_{ij}$'s and VRH relations. The equilibrium Pugh ratio for LDA, PBEsol, and PBE using SOC is ~ 0.22, ~ 0.24, and ~ 0.25, respectively, whereas without SOC, they are ~ 0.24, ~ 0.26, and ~ 0.28, respectively. These values are below the critical Pugh ratio of 0.57 (> for brittle and < for ductile), indicating ductile behaviour. Furthermore, the decrease in the Pugh ratio with increasing pressure signifies an enhancement in the ductility. The experimental [51] Pugh ratio agrees with our PBEsol$^{FR}$ result, whereas the reported [13] PBEsol with SOC values are significantly higher and are very close to our PBEsol$^{SR}$ values.

**CONCLUSION:**

The role of SOC in calculating the thermodynamical properties of lead has been studied comprehensively by comparing fully and scalar relativistic US-PPs. Additionally, the effect of different exchange correlational functionals coupled with FR and SR PPs is explored. We considered both the contribution of phonon and electronic excitation in calculating the Helmholtz free energy. Our findings show that the electronic excitation contribution in all studies is negligible. This result is expected as the $5d$ states are low in energy. However, incorporating $5d$ states in the valence is necessary since it improves the lattice constant [8]. The comparison with various experimental results and different theoretical reports is in reasonable agreement with the present study. We observe that the maximum difference in the equilibrium lattice constant and bulk modulus with and without SOC of about ~ 0.7 % and 22 % occurs for PBE functional. For the volume thermal expansion coefficient, the PBE$^{SR}$ gives a good result at low temperatures, and with the inclusion of SOC interaction, the thermal expansion coefficient increases significantly. The LDA$^{FR}$ and PBEsol$^{FR}$ $C_P$ follow the experiment reasonably well at low temperatures, whereas at high temperatures, the differences are significant. The experimental bulk modulus agrees with LDA$^{FR}$ and PBEsol$^{SR}$.

Analysis of the phonon dispersion at 100 K indicates that the introduction of SOC is crucial for accurately capturing the softening of the T branch at **X**. This result is in accordance with the previous theoretical reports. Furthermore, this effect of SOC at **X** is also visible in the mode-Grüneisen parameter $\gamma_{\mathbf{q}\eta}$, where the magnitude of $\gamma_{\mathbf{q}\eta}$ is significantly higher for the FR case and almost double at **X**. In contrast, at 300 K, we notice that including SOC substantially decreases the overall phonon frequencies with respect to the experiment, and in this case, the scalar relativistic functional performs better, except for the T branch at **X**.

The thermodynamic average Grüneisen parameter increases with temperature when SOC is considered, while it remains almost constant for the SR case. Our results for thermal expansion coefficient, isobaric heat capacity, and phonon dispersion for both SR and FR conditions are similar for LDA and PBEsol. The contribution $((FR - SR)/FR)$ of the SOC effect on different thermodynamic properties discussed in supplementary S1 for PBEsol shows that with increasing pressure, the contributions of SOC in the thermodynamic properties decrease but do not vanish. This is in agreement with Smirnov [13]. The experimental elastic constant-coefficient and shear modulus are close to LDA[FR] and PBEsol[FR]. In contrast, PBE[SR] agrees well with Debye temperature, and polycrystalline longitudinal and transverse sound velocities agree with the previous studies. The pressure-dependent elastic constant and Pugh ratio at 0 K indicate the enhancement in the ductility of lead prevails with increasing pressure.

## ACKNOWLEDGMENT

Computational facilities were provided by SISSA through its Linux Cluster and ITCS and SISSA-CINECA 2021-2024 agreement. This work has been supported by the Italian MUR through the National Centre for HPC, Big Data, and Quantum Computing (grant No. CN000 00013).

## DATA AVAILABILITY STATEMENT :

All data that support the findings of the present study are included within the article.

## REFERENCES:

[1] R. F. Tylecote, *The Behaviour of Lead as a Corrosion Resistant Medium Undersea and in Soils*, J. Archaeol. Sci. **10**, 397 (1983).

[2] S. B. Lyon, *Corrosion of Lead and Its Alloys*, in *Shreir's Corrosion* (Elsevier, 2010), pp. 2053–2067.

[3] D. Rezaei-Ochbelagh and S. Azimkhani, *Investigation of Gamma-Ray Shielding Properties of Concrete Containing Different Percentages of Lead*, Appl. Radiat. Isot. **70**, 2282 (2012).

[4] L. Kleinman, *Relativistic Norm-Conserving Pseudopotential*, Phys. Rev. B **21**, 2630 (1980).

[5] A. Dal Corso and A. M. Conte, *Spin-Orbit Coupling with Ultrasoft Pseudopotentials: Application to Au and Pt*, Phys. Rev. B **71**, 115106 (2005).

[6] M. J. Verstraete, M. Torrent, F. Jollet, G. Zérah, and X. Gonze, *Density Functional Perturbation Theory with Spin-Orbit Coupling: Phonon Band Structure of Lead*, Phys. Rev. B **78**, 045119 (2008).

[7] B. N. Brockhouse, T. Arase, G. Caglioti, K. R. Rao, and A. D. B. Woods, *Crystal Dynamics of Lead. I. Dispersion Curves at 100°K*, Phys. Rev. **128**, 1099 (1962).

[8] A. Dal Corso, *Ab Initio Phonon Dispersions of Face Centered Cubic Pb: Effects of Spin–Orbit Coupling*, J. Phys. Condens. Matter **20**, 445202 (2008).

[9] S. de Gironcoli, *Lattice Dynamics of Metals from Density-Functional Perturbation Theory*, Phys. Rev. B **51**, 6773 (1995).

[10] X. M. Chen and A. W. Overhauser, *Effect of Spin-Density Waves on the Lattice Dynamics of Lead*, Phys. Rev. B **39**, 10570 (1989).

[11] A. Kuznetsov, V. Dmitriev, L. Dubrovinsky, V. Prakapenka, and H.-P. Weber, *FCC–HCP Phase Boundary in Lead*, Solid State Commun. **122**, 125 (2002).

[12] H. K. Mao, Y. Wu, J. F. Shu, J. Z. Hu, R. J. Hemley, and D. E. Cox, *High-Pressure Phase Transition and Equation of State of Lead to 238 GPa*, Solid State Commun. **74**, 1027 (1990).

[13] N. A. Smirnov, *Effect of Spin-Orbit Interactions on the Structural Stability, Thermodynamic Properties, and Transport Properties of Lead under Pressure*, Phys. Rev. B **97**, 094114 (2018).

[14] B. Grabowski, T. Hickel, and J. Neugebauer, *Ab Initio Study of the Thermodynamic Properties of Nonmagnetic Elementary Fcc Metals: Exchange-Correlation-Related Error Bars and Chemical Trends*, Phys. Rev. B **76**, 024309 (2007).

[15] J. P. Perdew, A. Ruzsinszky, G. I. Csonka, O. A. Vydrov, G. E. Scuseria, L. A. Constantin, X. Zhou, and K. Burke, *Restoring the Density-Gradient Expansion for Exchange in Solids and Surfaces*, Phys. Rev. Lett. **100**, 136406 (2008).

[16] B. Thakur, X. Gong, and A. Dal Corso, *Ab Initio Thermodynamic Properties of Iridium: A High-Pressure and High-Temperature Study*, Comput. Mater. Sci. **234**, 112797 (2024).

[17] B. Thakur, X. Gong, and A. Dal Corso, *Thermodynamic Properties of Rhodium—A First Principle Study*, AIP Adv. **14**, 045229 (2024).

[18] L. Kývala and D. Legut, *Lattice Dynamics and Thermal Properties of Thorium Metal and Thorium Monocarbide*, Phys. Rev. B **101**, 075117 (2020).

[19] A. Dal Corso, *The Thermo_pw Code Can Be Downloaded from the Web Page*, https://dalcorso.github.io/thermo_pw/.

[20] P. Giannozzi et al., *QUANTUM ESPRESSO: A Modular and Open-Source Software Project for Quantum Simulations of Materials*, J. Phys. Condens. Matter **21**, 395502 (2009).

[21] P. Giannozzi et al., *Advanced Capabilities for Materials Modelling with Quantum ESPRESSO*, J. Phys. Condens. Matter **29**, 465901 (2017).

[22] D. Vanderbilt, *Soft Self-Consistent Pseudopotentials in a Generalized Eigenvalue Formalism*, Phys. Rev. B **41**, 7892 (1990).

[23] A. Dal Corso, *The Pslibrary Pseudopotential Library*, https://github.com/dalcorso/pslibrary/.

[24] A. Dal Corso, *Pseudopotentials Periodic Table: From H to Pu*, Comput. Mater. Sci. **95**, 337 (2014).

[25] A. M. Rappe, K. M. Rabe, E. Kaxiras, and J. D. Joannopoulos, *Optimized Pseudopotentials*, Phys. Rev. B **41**, 1227 (1990).

[26] A. Dal Corso, *Density Functional Perturbation Theory for Lattice Dynamics with Fully Relativistic Ultrasoft Pseudopotentials: Application to Fcc-Pt and Fcc-Au*, Phys. Rev. B **76**, 054308 (2007).

[27] J. P. Perdew and A. Zunger, *Self-Interaction Correction to Density-Functional Approximations for Many-Electron Systems*, Phys. Rev. B **23**, 5048 (1981).

[28] J. P. Perdew, K. Burke, and M. Ernzerhof, *Generalized Gradient Approximation Made Simple*, Phys. Rev. Lett. **77**, 3865 (1996).

[29] S. G. Louie, S. Froyen, and M. L. Cohen, *Nonlinear Ionic Pseudopotentials in Spin-Density-Functional Calculations*, Phys. Rev. B **26**, 1738 (1982).

[30] H. J. Monkhorst and J. D. Pack, *Special Points for Brillouin-Zone Integrations*, Phys. Rev. B **13**, 5188 (1976).

[31] M. Methfessel and A. T. Paxton, *High-Precision Sampling for Brillouin-Zone Integration in Metals*, Phys. Rev. B **40**, 3616 (1989).

[32] S. Baroni, S. de Gironcoli, A. Dal Corso, and P. Giannozzi, *Phonons and Related Crystal Properties from Density-Functional Perturbation Theory*, Rev. Mod. Phys. **73**, 515 (2001).

[33] C. Malica and A. Dal Corso, *Temperature Dependent Elastic Constants and Thermodynamic Properties of BAs: An Ab Initio Investigation*, J. Appl. Phys. **127**, 245103 (2020).

[34] C. Malica and A. Dal Corso, *Quasi-Harmonic Thermoelasticity of Palladium, Platinum, Copper, and Gold from First Principles*, J. Phys. Condens. Matter **33**, 475901 (2021).

[35] A. Dal Corso, *Elastic Constants of Beryllium: A First-Principles Investigation*, J. Phys. Condens. Matter **28**, 075401 (2016).

[36] X. Gong and A. Dal Corso, *Ab Initio Quasi-Harmonic Thermoelasticity of Molybdenum at High Temperature and Pressure*, J. Chem. Phys. **160**, 244703 (2024).

[37] X. Gong and A. Dal Corso, *Pressure and Temperature Dependent Ab-Initio Quasi-Harmonic Thermoelastic Properties of Tungsten*, J. Phys. Condens. Matter **36**, 285702 (2024).

[38] R. Hill, *The Elastic Behaviour of a Crystalline Aggregate*, Proc. Phys. Soc. Sect. A **65**, 349 (1952).

[39] T. Strässle, S. Klotz, K. Kunc, V. Pomjakushin, and J. S. White, *Equation of State of Lead from High-Pressure Neutron Diffraction up to 8.9 GPa and Its Implication for the NaCl Pressure Scale*, Phys. Rev. B **90**, 014101 (2014).

[40] P. Söderlind and D. Young, *Assessing Density-Functional Theory for Equation-Of-State*, Computation **6**, 13 (2018).

[41] Y. S. Touloukian, R. K. Kirby, R. E. Taylor, and P. D. Desai, *Thermal Expansion - Metallic Elements and Alloys*, Thermophys. Prop. Matter-the TPRC Data Ser. **12**, 178 (1975).

[42] D. E. Gray, *American Institute of Physics Handbook* (McGraw-Hill Book Company, New York, 1957).

[43] N. A. Smirnov, *Ab Initio Calculations of Structural Stability, Thermodynamic and Elastic Properties of Ni, Pd, Rh, and Ir at High Pressures*, J. Appl. Phys. **134**, 025901 (2023).

[44] D. C. Wallace, *Thermodynamics of Crystals* (Dover Publications, 1998).

[45] J. H. Rose, J. R. Smith, F. Guinea, and J. Ferrante, *Universal Features of the Equation of State of Metals*, Phys. Rev. B **29**, 2963 (1984).

[46] P. Haas, F. Tran, and P. Blaha, *Calculation of the Lattice Constant of Solids with Semilocal Functionals*, Phys. Rev. B **79**, 085104 (2009).

[47] A. R. Stokes and A. J. C. Wilson, *The Thermal Expansion of Lead from 0 c. to 320 C*, Proc. Phys. Soc. **53**, 658 (1941).

[48] S. V. Stankus and R. A. Khairulin, *The Density of Alloys of Tin—Lead System in the Solid and Liquid States*, High Temp. **44**, 389 (2006).

[49] T. Rubin, H. L. Johnston, and H. W. Altman, *The Thermal Expansion of Lead*, J. Phys. Chem. **66**, 266 (1962).

[50] J. W. Arblaster, *Thermodynamic Properties of Lead*, Calphad **39**, 47 (2012).

[51] D. L. Waldorf and G. A. Alers, *Low-Temperature Elastic Moduli of Lead*, J. Appl. Phys. **33**, 3266 (1962).

[52] C. L. Vold, M. E. Glicksman, E. W. Kammer, and L. C. Cardinal, *The Elastic Constants for Single-Crystal Lead and Indium from Room Temperature to the Melting Point*, J. Phys. Chem. Solids **38**, 157 (1977).

[53] G. Cordoba and C. R. Brooks, *The Heat Capacity of Lead from 300 to 850 °K Conversion of Cp to Cv for Solid Lead*, Phys. Status Solidi **11**, 749 (1972).

[54] N. V. Kozyrev and V. V. Gordeev, *Thermodynamic Characterization and Equation of State for Solid and Liquid Lead*, Metals (Basel). **12**, 16 (2021).

[55] R. Stedman, L. Almqvist, G. Nilsson, and G. Raunio, *Dispersion Relations for Phonons in Lead at 80 and 300°K*, Phys. Rev. **162**, 545 (1967).

[56] O. P. Gupta, *Thermal Properties of Lead*, Phys. B+C **122**, 236 (1983).

[57] R. A. MacDonald and W. M. MacDonald, *Thermodynamic Properties of Fcc Metals at High Temperatures*, Phys. Rev. B **24**, 1715 (1981).

# Supplementary Data

**Thermodynamic properties of fcc lead: A scalar and fully relativistic first principle study**

Balaram Thakur[1*], Xuejun Gong[1,2], and Andrea Dal Corso[1,2]

[1]International School for Advanced Studies (SISSA), Via Bonomea 265, 34136 Trieste, Italy.

[2]CNR-IOM, Via Bonomea 265, 34136 Trieste, Italy.

*Corresponding author: bthakur@sissa.it

Email:  Balaram Thakur (bthakur@sissa.it), Xuejun Gong (xgong@sissa.it), Andrea Dal Corso (dalcorso@sissa.it)

## S1: Contribution of spin-orbit coupling (SOC) in the thermodynamic properties of lead:

In the main article, we reported the effect of fully relativistic (FR) and scalar relativistic (SR) pseudopotential and the role of LDA, PBEsol, and PBE exchange-correlational functional on the thermodynamic properties of fcc-lead. Here we illustrate the contribution $((FR-SR)/FR)$ of SOC in volume thermal expansion coefficient, heat capacity, bulk modulus, and average Grüneisen parameter. For this, we determined the difference in the respective thermodynamic properties obtained using with and without SOC, as a function of temperature and pressure, and for PBEsol functional is discussed below:

Volume thermal expansion coefficient (β):

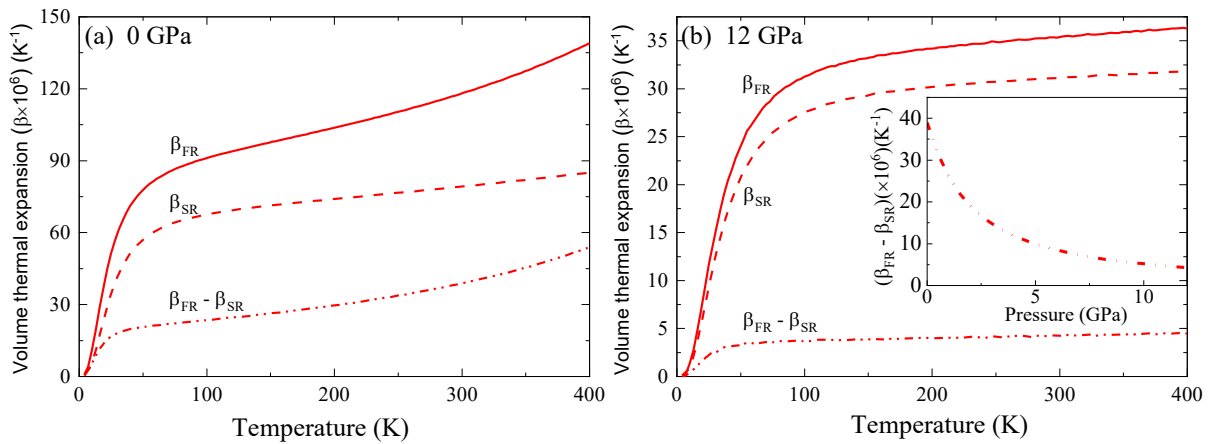

**Figure S1A**: The temperature-dependent volume thermal expansion coefficient (β) for PBEsol using fully relativistic PP ($β_{FR}$), scalar relativistic PP ($β_{SR}$), and their difference ($β_{FR}$-$β_{SR}$) at (a) 0 GPa and (b) 12 GPa. The inset of (b) shows the pressure dependence of $β_{FR}$-$β_{SR}$ at 301 K for PBEsol.

Heat Capacity:

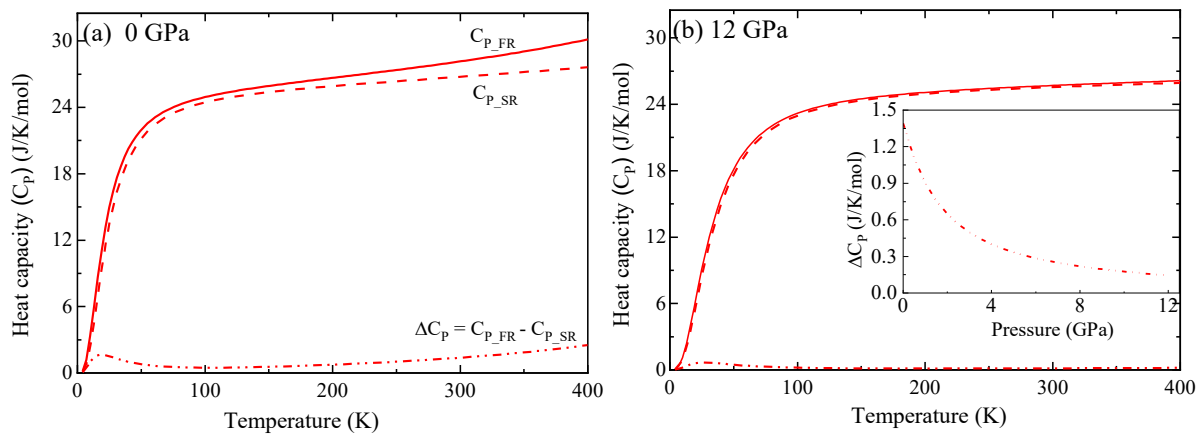

**Figure S1B**: The temperature-dependent isobaric heat capacity ($C_P$) for PBEsol using fully relativistic PP ($C_{P\_FR}$), scalar relativistic PP ($C_{P\_SR}$), and their difference ($C_{P\_FR}$ - $C_{P\_SR}$) at (a) 0 GPa and (b) 12 GPa. The inset of (b) shows the pressure dependence of $C_{P\_FR}$ - $C_{P\_SR}$ at 301 K for PBEsol. The enhancement in the $C_{P\_FR}$ - $C_{P\_SR}$ in (a) with increasing temperature is due to the contribution of β as shown in Figure S1A.

Thermodynamic average Grüneisen parameter (γ):

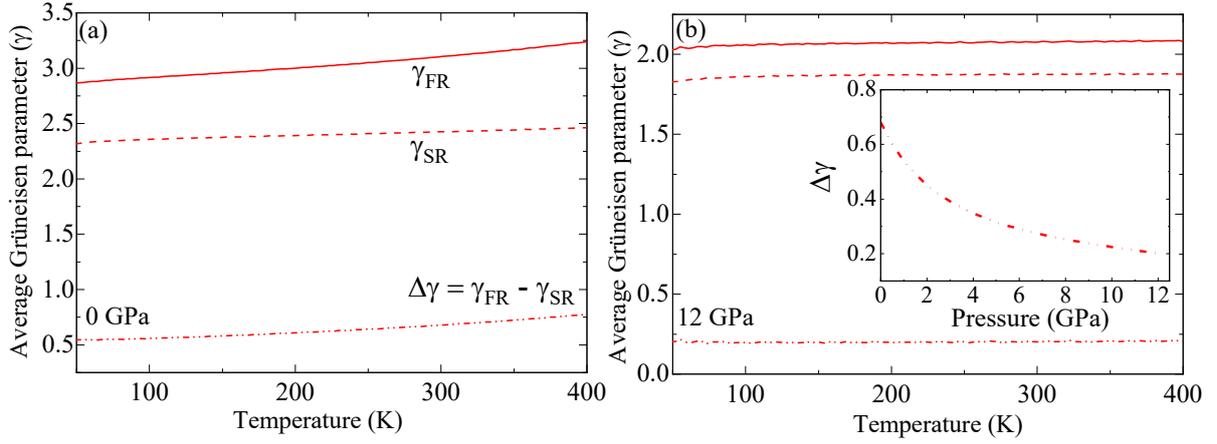

**Figure S1C**: The temperature-dependent average Grüneisen parameter (γ) for PBEsol using fully relativistic PP ($\gamma_{FR}$), scalar relativistic PP ($\gamma_{SR}$), and their difference ($\gamma_{FR}$- $\gamma_{SR}$) at (a) 0 GPa and (b) 12 GPa. The inset of (b) shows the pressure dependence of $\gamma_{FR}$- $\gamma_{SR}$ at 301 K for PBEsol.

Bulk Modulus:

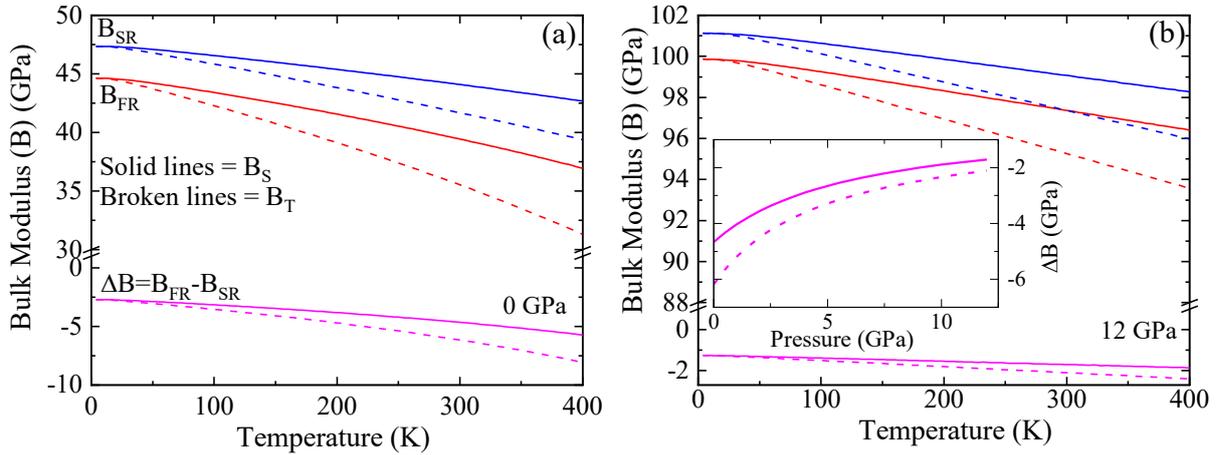

**Figure S1D**: The temperature-dependent bulk modulus (B) for PBEsol using fully relativistic PP ($B_{FR}$), scalar relativistic PP ($B_{SR}$), and their difference ($B_{FR}$- $B_{SR}$) at (a) 0 GPa and (b) 12 GPa. The inset of (b) shows the pressure dependence of $B_{FR}$- $B_{SR}$ at 301 K for PBEsol. The solid and broken lines in (a) and (b) represent the isoenthalpic ($B_S$) and isoentropic ($B_T$) bulk modulus, respectively.

From Fig. S1A to Fig.S1D, we observed that the contribution of the SOC effect is significant in the volume thermal expansion coefficient where the $\beta_{FR}$-$\beta_{SR}$ increases with temperature. However, the SOC effect was significantly suppressed with increasing pressure, as shown in the inset of Figures S1A(b) to S1D(b). We found that at 400 K, the SOC contribution decreases with increasing the pressure from 0 GPa to 12 GPa. For example, the thermal expansion coefficient changes from 39 % to 12 %, the isobaric heat capacity from 8 % to less than 1%, the average Grüneisen parameter from 24 % to 10 %, the isothermal bulk modulus from 26 % to 3 %, and the isoenthalpic bulk modulus from 16 % to 2 %. Our study is in agreement with Smirnov [1], where the author found that the contribution of SOC is reduced but does not vanish under higher pressure.

**References:**

[1]  N.A. Smirnov, Effect of spin-orbit interactions on the structural stability, thermodynamic properties, and transport properties of lead under pressure, Phys. Rev. B. 97 (2018) 094114. https://doi.org/10.1103/PhysRevB.97.094114.